\documentclass{aa}

\usepackage{amsmath}
\usepackage{graphicx} 
\usepackage[dvipsnames]{xcolor}
\usepackage{siunitx}
\usepackage{lineno}
\usepackage{setspace}
\usepackage{txfonts}
\usepackage{esdiff}
\usepackage{bm}
\usepackage{natbib}
\usepackage{hyperref}
\hypersetup{
    colorlinks=true,
    linkcolor=blue,
    citecolor=blue,
    urlcolor=magenta}

\begin{document}

\title{From core to envelope: Revealing the deep dynamics of stars with two convective zones}

\titlerunning{}

\author{S.N.~Breton\inst{1}
    \and 
    A.S.~Brun\inst{2}
    \and
    R.A.~García\inst{2}}

    \institute{INAF – Osservatorio Astrofisico di Catania, Via S. Sofia, 78, 95123 Catania, Italy.
    \\
    \email{sylvain.breton@inaf.it} 
    \and
    Universit\'e Paris-Saclay, Universit\'e Paris Cit\'e, CEA, CNRS, AIM, 91191, Gif-sur-Yvette, France.}

\date{}

\abstract{On the Hertzsprung-Russell diagram, F-type solar pulsators connect the Sun to intermediate-mass stars located on the instability strip. {With respect to lower-mass stars, they are structurally peculiar in the sense that} they are constituted of three distinct dynamical layers: a small convective core, a deep radiative interior, and a shallow convective envelope. Current asteroseismic techniques only provide limited information on the interior dynamics of these stars. Indeed only gravity modes ($g$ modes), for which unambiguous characterisation is lacking, are able to probe the deep stellar layers. A better understanding of the excitation and behaviour in F-type solar pulsators is therefore necessary in order to consider their detection.
In this work, we simulate the global stellar structure of an F-type star (core, radiative interior, envelope) for the first time. We show that the contribution of the core strongly affects the spectrum of excited $g$ modes, with low-order high-degree modes unable to form due to their interaction with the turbulent convection of the core. Finally, by computing the disc-integrated signature of the modes, we are able to demonstrate that they preserve their integrity up to the top of the convective envelope, which is a strong argument in favour of their detectability with space-borne photometry.}


\maketitle

\section{Introduction}

With their {peculiar structural properties, F-type stars populate the region between the Sun and intermediate-mass stars \citep[see e.g. Fig.~1 from][]{Breton2023}}. Their radiative interior is surrounded both by a shallow convective envelope and a small convective core, which makes them fascinating test-beds to explore the non-linear excitation of internal waves by decoupled convective sources. When they are able to form standing modes, these internal waves give rise to stellar oscillations \citep[e.g.][]{Unno1989,Aerts2010}. In particular, direct constraints on the core properties and on the stellar deep dynamics can only be obtained through the observation of gravity modes \citep[$g$ modes, e.g.][]{Christensen-Dalsgaard1985Sci,Garcia2007}, which are the standing modes of oscillation related to the propagation of internal gravity waves (IGWs) in the stellar interior. However, due to their low surface amplitude when a convective envelope is present, the detectability of such $g$ modes is unclear \citep{Belkacem2009,Belkacem2022,Breton2023} below the $\gamma$~Dor instability strip \citep[e.g.][]{Kaye1999,Warner2003,Dupret2004}. 

F-type main-sequence stars with an effective temperature below 6500~K exhibit stochastically excited acoustic modes of oscillation \citep[$p$ modes, e.g.][]{Belkacem2008,Chaplin2011,Mathur2014,Lund2017} and are therefore still considered solar-like from this point of view. They are of paramount importance for our understanding of stellar objects below 1.5~$\rm M_\odot$, since a significant part of the upcoming Planetary Transit and Oscillation of Stars mission \citep[PLATO,][]{Rauer2025} core sample will be made up of such main-sequence F-type solar-like pulsators \citep{Montalto2021,Goupil2024}. Although $p$-mode asteroseismology enables us to put some constraints on the extent of the convective core of F-type stars, this requires the implementation of model-dependent fitting of observational frequencies on a stellar evolutionary track \citep[e.g.][]{Deheuvels2016}, or the application of complex inversion techniques \citep[e.g.][]{Betrisey2022}. Moreover, $p$ modes alone are unable to provide constraints on the rotational dynamics of the deep radiative layers \citep[e.g.][]{Thompson1996Sci,Thompson2003,Garcia2007,Benomar2015}. On the contrary, the core extent is directly encoded in the $g$-mode asymptotic period spacing \citep{Tassoul1980}, and $g$ modes are extremely sensitive to the rotation of such regions \citep{Deheuvels2012,Li2020}. The main issue for the application of $g$-mode asteroseismology on main-sequence solar-type stars stems from the fact that, unlike $\gamma$~Dor stars that exhibit $g$ modes with large amplitudes due to the existence of non-adiabatic destabilising mechanisms \citep[e.g.][]{Dupret2005}, $g$ modes in solar-type stars are stable and can only be excited by the stochastic energy input from turbulent convective motions \citep[e.g.][]{Kumar1996,Lecoanet2013,Pincon2016,Augustson2020}. Previous 2D and 3D hydrodynamical simulations in polar and spherical configurations demonstrated that IGWs are more efficiently excited  in F-type stars \citep{Breton2022simuFstars} than in the Sun \citep[e.g.][]{Rogers2005,Brun2011,Alvan2014,Alvan2015}.

In main-sequence solar-type stars, $g$ modes are low-frequency ($\nu \lesssim 500 \, \mu$Hz) modes with an absolute radial order, $n$, that increases with decreasing frequency. In F-type stars, low-order high-frequency modes are in principle able to mix with low-order $p$ modes \citep{Breton2023} to form mixed modes, while high-order low-frequency $g$ modes will become gravito-inertial modes under the effect of the Coriolis force \citep{Lee1997,Townsend2003,VanReeth2015,VanReeth2018}, with possible additional couplings between the eigenmodes of the core, the radiative interior, and the envelope \citep[][]{Ouazzani2020,Tokuno2022,Barrault2025,Breton2026}. Unambiguously observing and characterising $g$ modes in late F-type stars, under any of their manifestations, would bring new, strong structural constraints on stellar evolution processes in solar-type stars. Moreover, measuring their rotational splittings would open a direct window onto the rotation of the deep radiative layers. Both of these aspects would represent an invaluable asset for our understanding of stellar structure and evolution.   

In this work, we present the first 3D simulations of a solar-type stellar interior including both a convective core and a convective envelope \citep[see][for a two-convective zones simulation in the case of an O-type star]{Pathak2026}.
We show that the $g$ modes encode the structural extent of the core in their properties. The modes maintain their coherence up to the highest regions of the convective envelope. With respect to models that only account for envelope excitation, we find that the core plays a crucial role by filtering out high-frequency, high-degree modes. By constructing observational time series from our simulations, we are able to unambiguously identify the signature of the modes at the top of the convective envelope. This should justify and encourage future efforts for their identification and characterisation in observational time series. The structure of the article is as follows. In Sect.~\ref{sec:simulation_setups}, we present the anelastic numerical setup of our simulations. In Sect.~\ref{sec:mode_power_spectrum}, we compute the power spectrum of the $g$ modes excited in the simulation in order to analyse the different behaviour exhibited by the setups depending on the mode excitation source. Section~\ref{sec:mode_detectability} is dedicated to the computation of pseudo-observational time series and the discussion of the mode detectability. Finally, we draw the conclusions of our work in Sect.~\ref{sec:conclusion}.  

\section{Simulation setups \label{sec:simulation_setups}}

\begin{figure*}[ht!]
    \centering
    \includegraphics[width=0.99\linewidth]{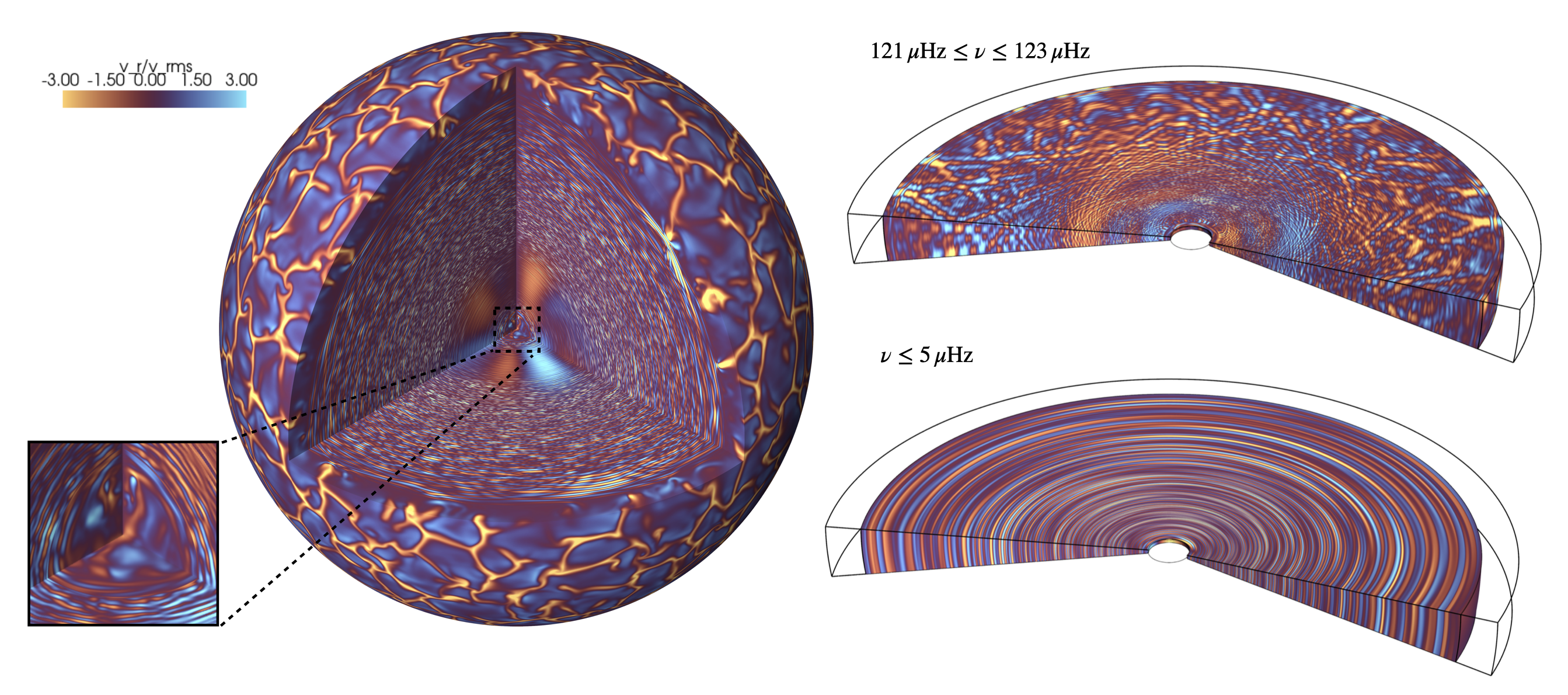}
    \caption{\textit{Left:} Volume display of the radial velocity field in the Full model, normalised by the shell root-mean square value, $\varv_r / \varv_{\rm rms}$. The inset on the left displays a zoom on the convective core. \textit{Top right:} Volume display of $\varv_r / \varv_{\rm rms}$ after filtering the signal to isolate the signal at $121 \, \mu\mathrm{Hz} \leq \nu \leq 123 \, \mu\mathrm{Hz}$. \textit{Bottom right:} Same as above, but isolating the signal at $\nu \leq 5 \, \mu\mathrm{Hz}$.}
    \label{fig:volume_display}
\end{figure*}

\begin{figure*}[ht!]
    \centering
    \includegraphics[width=0.8\linewidth]{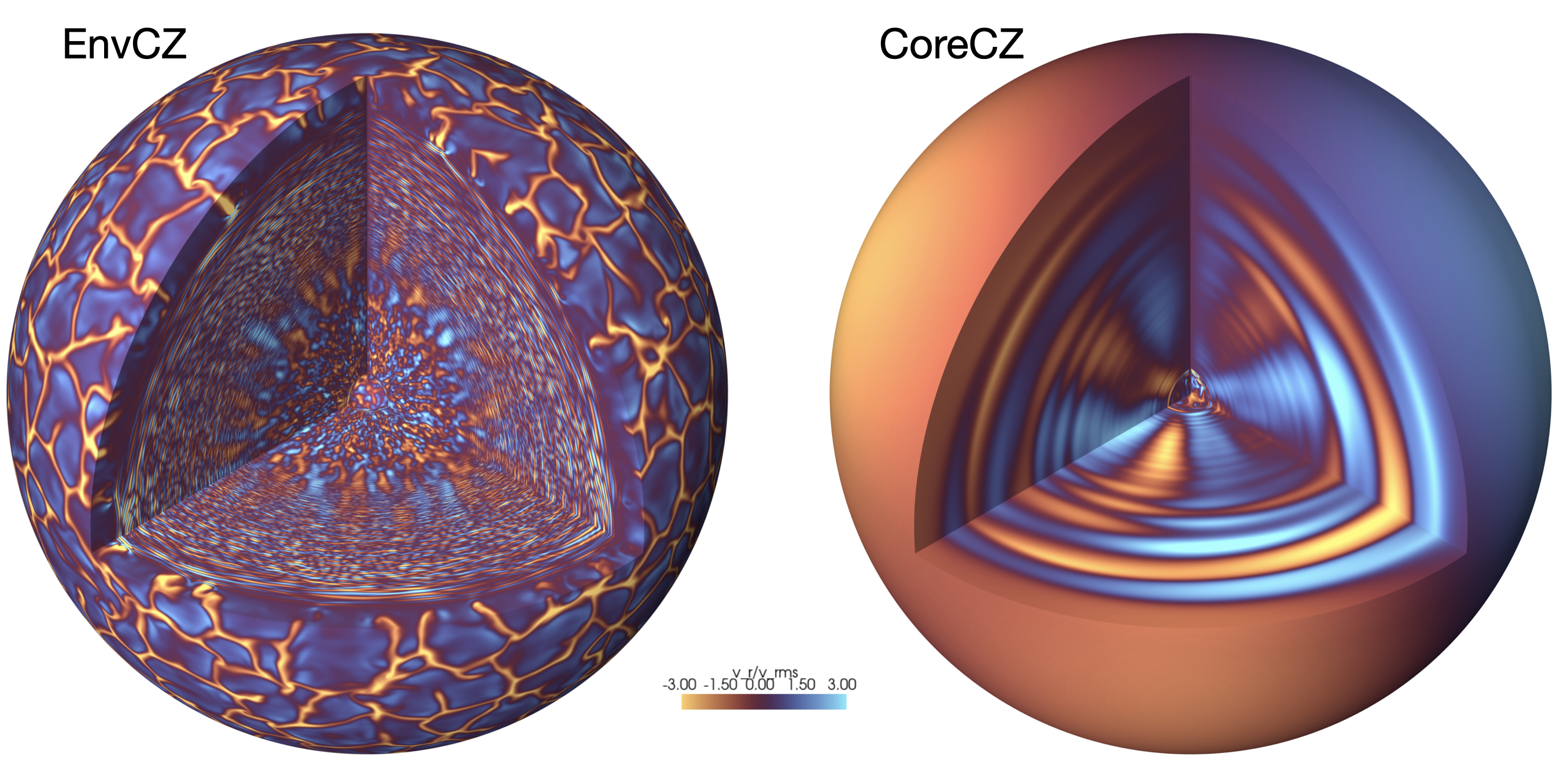}
    \caption{Volume display of the radial velocity field in the EnvCZ model (\textit{left}) and the CoreCZ model (\textit{right}), normalised by the shell root-mean square value, $\varv_r / \varv_{\rm rms}$.}
    \label{fig:auxiliary_models}
\end{figure*}

In order to assess the importance of the convective core in the internal wave dynamics of an F-type stars, we developed and we analysed three 1.3~$\rm M_\odot$ spherical models with the Anelastic Spherical Harmonics code \citep[ASH,][]{Clune1999,Brun2004}. The first one includes the full extent of the stellar structure (convective envelope surrounding the radiative interior and the convective core; the model is named 'Full' hereafter). The second one only includes the convective envelope and the radiative interior (model named  'EnvCZ' hereafter), and the third one only includes the radiative interior and the convective core (model named  'CoreCZ' hereafter). 
We note that, despite their similar configuration, the background structure of the EnvCZ model slightly differs from the model presented in \citet{Breton2022simuFstars}.
Each simulation is initialised with small perturbations in the unstable convective layer to allow for the growth of convective instability, and evolved until it statistically reaches a relaxed steady state. More details on the numerical setup used for the the simulations are provided in Appendix~\ref{sec:numerical_setup}. 

A volume representation of the radial velocity field, $\varv_r$, for the Full model is shown in Fig.~\ref{fig:volume_display}. To enable a visualisation of both the convective structure and the IGW pattern, $\varv_r$ is normalised by the root-mean-square value, $\varv_{\rm rms}$, computed on the radial shell to which each cell belongs. In the convective envelope, large-scale upwards and downwards flows are visible. In particular, downwards motions can be seen interacting with the top of the radiative interior through plumes penetrating the stably stratified layers and exciting the IGWs. Given the more limited extent of the core, the typical scale of the convective motions close to the centre of the star is smaller than for the envelope. As shown in Fig.~\ref{fig:v_rms_comparison} and expected from a mixing-length argument \citep{BohmVitense1958}, the convective motions are also slower in the core than in the envelope. In the upper layers of the radiative interior, the signal is constituted by the superposition of progressive IGWs and standing $g$ modes of all degrees. Close to the convective core, the wave signal is dominated by high-amplitude, high-frequency $g$ modes. For comparison with the Full model,  the volume display of the EnvCZ and CoreCZ models are shown in Fig.~\ref{fig:auxiliary_models}. We note that, due to the absence of the convective core, the superposition of mode patterns in the lower half of the radiative interior differs between the Full and the EnvCZ cases. Finally, the wave structure we distinguish in the CoreCZ case has a large scale even close to the top of the radiative interior, which suggests that the signal is dominated by low-order modes with few radial nodes.

\begin{figure}[ht!]
    \centering
    \includegraphics[width=0.99\linewidth]{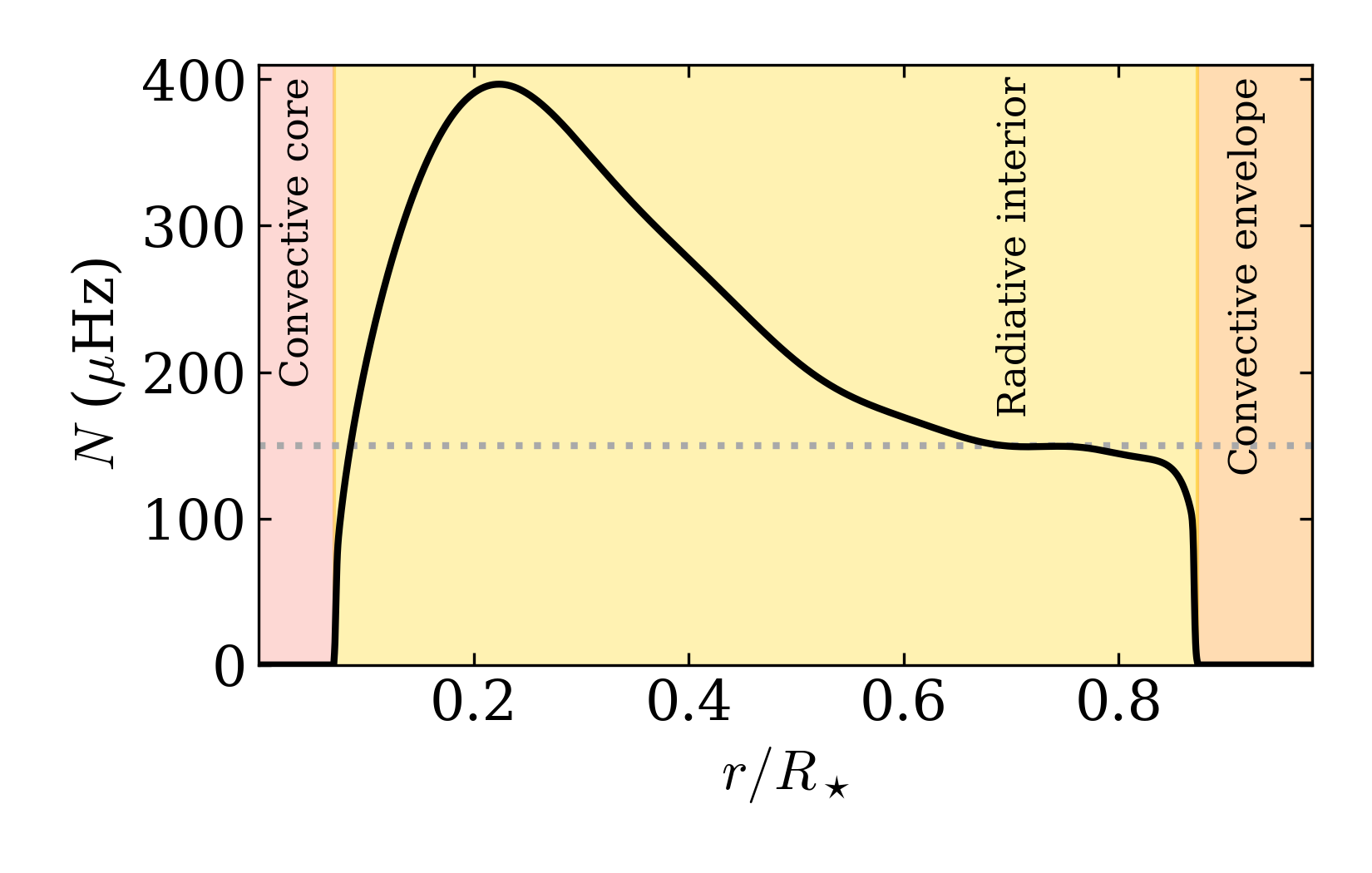}
    \caption{Profile of the Brunt-Väisäla frequency of the reference stellar structure, $N$. The extent of the convective core, the radiative interior, and the convective envelope are indicated in colour. The value of $N$ was set to zero in the convective regions where $N^2<0$. The frequency limit where intermediate- and high-degree modes are unable to form in the Full model while visible in the EnvCZ model is indicated with the dotted grey line.}
    \label{fig:brunt_vaisala}
\end{figure}

In Fig.~\ref{fig:brunt_vaisala} we show the profile of the Brunt-Väisäla frequency, $N$, of the reference stellar structure. We have $N=0$ in the convective layer and $N>0$ in the radiative layers. Given that $N$ imposes the maximal frequency of IGWs propagating in the local medium, waves in the radiative interior are expected to have frequencies up to 400~$\mu$Hz. As the ray trajectory of an IGW is frequency-dependent \citep{Gough1993}, it is possible to filter the temporal signal to disentangle the signature of waves at distinct frequencies \citep{Alvan2015}. We illustrate this in the right panel of Fig.~\ref{fig:volume_display} where we show the result of this filtering in the radiative interior for a frequency of $121 \, \mu\mathrm{Hz} \leq \nu \leq 123 \, \mu\mathrm{Hz}$ and $\nu \leq 5 \, \mu\mathrm{Hz}$. In the former case, the ray travels from inner caustic to outer caustic, with a reflection geometry that creates the typical rosetta pattern of IGWs in this range of frequency. In the latter case, at very low frequency, the rays converge towards the centre in a spiral pattern, with winding angles so low that they appear as concentric circles.

\section{Mode power spectrum \label{sec:mode_power_spectrum}}

\begin{figure*}[ht!]
    \centering
    \includegraphics[width=0.99 \textwidth]{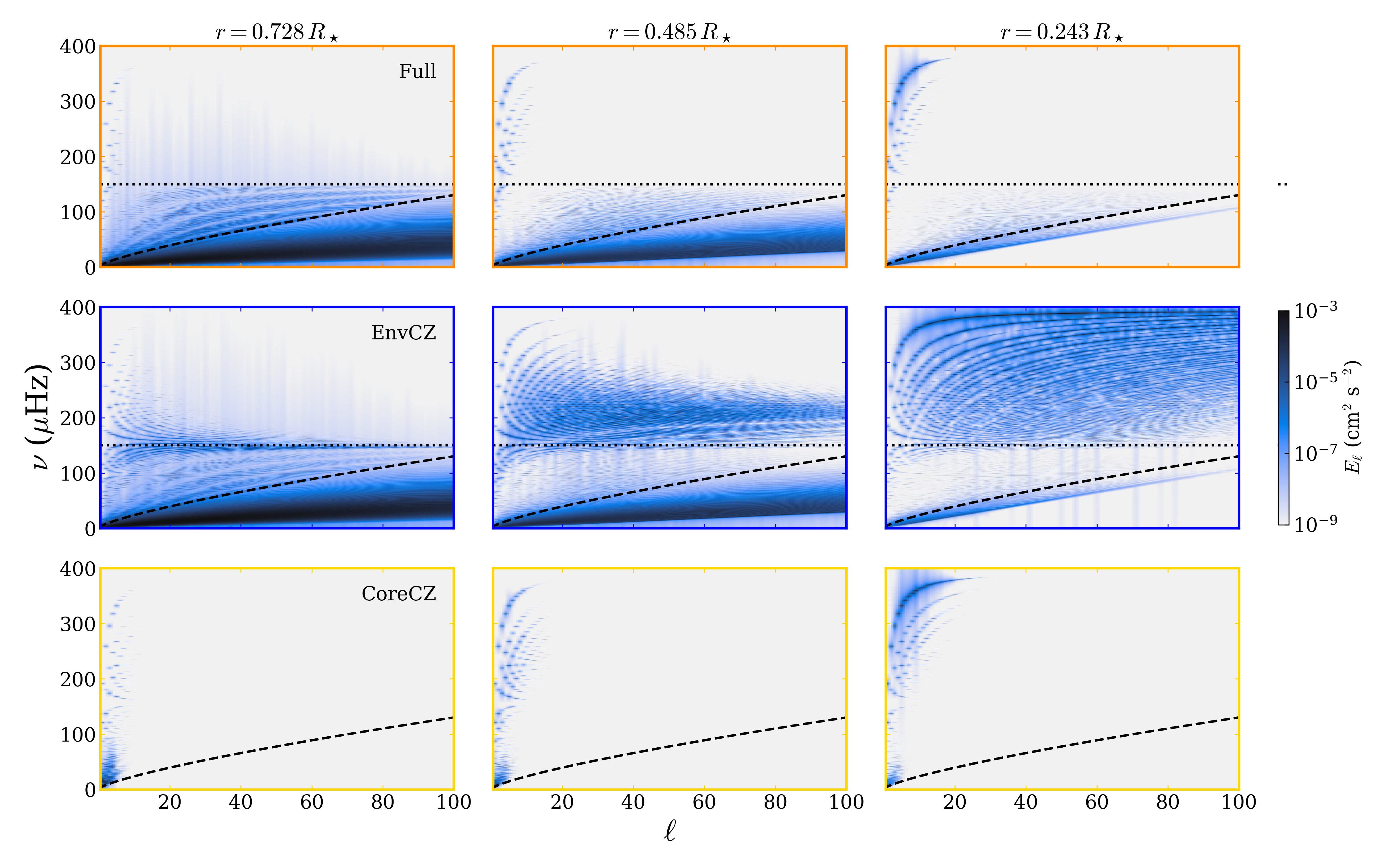}
    \caption{Power spectra $E_\ell$ at depths $r = [0.243, 0.485, 0.728]$~$R_\star$ for the Full model (first row), the EnvCZ model (second row), and the CoreCZ model (third row). The degree-dependent cut-off frequency between the standing modes and the progressive waves is shown on every panel with a dashed black line. In the first two rows, the limit frequency where intermediate- and high-degree modes are unable to form in the Full model while visible in the EnvCZ model is indicated with the dotted black line. {The border colour of the panels is different for each row in order to guide the eye (each colour corresponds to a model: orange, blue, and yellow for the Full, EnvCZ, and CoreCZ models, respectively).}}
\label{fig:spectrum_combined_nine_panels_radial_velocity}
\end{figure*}

Once the simulations are relaxed, we computed the power spectrum of the radial velocity at given depths for our different setups. We considered time series of lengths 104~days.   

\subsection{Power spectrum computation}

For each 3D model, the power spectrum was computed by constructing time series with the simulation outputs \citep{Breton2022simuFstars}. Considering the radial grid of the simulations, a sub-sample of spherical shells with radial coordinates, $r$, were selected. The corresponding $(\theta, \phi)$ velocity maps, with $\theta$ the latitude and $\phi$ the longitude, were output with a regular sampling $\mathrm{d}t = 600$~s, which corresponds to a Nyquist frequency of approximately 833~$\mu$Hz, which is larger than the maximal value of our structural $N$ profile. We computed the spherical harmonic transform of each spherical map in order to obtain the $\ell, m$ coefficients of the velocity field, with $\ell$ the angular degree and $m$ the azimuthal number.
{Given that these high-cadence time series represent a very large amount of data, we computed this transform up to $\ell = 100$, as a trade-off between preservation of the signal at a low-angular scale and data storage availability.}
We then applied a discrete Fourier transform (DFT). The transformation process can then be summarised as
\begin{equation}
    \varv_r (r, \theta, \phi, t) \rightarrow \tilde{\varv}_r (r, \ell, m, t) \rightarrow   \hat{\varv}_r (r, \ell, m, \nu) \; ,
\end{equation}
where the complex quantities $\tilde{\varv}_r$ and $\hat{\varv}_r$ are computed from the radial velocity $\varv_r$. Given the 104-day length of the time series, the frequency bin width is $\sim 0.11$~$\mu$Hz. To obtain the energy coefficients with respect of each angular degree, $E_\ell$, that we represent in Fig~\ref{fig:spectrum_combined_nine_panels_radial_velocity}, we computed the quadratic sum of the modulus of the $m$-components for each $\ell:$
\begin{equation}
    E_\ell (r) = \sum_{m=-\ell}^{m=\ell} |\hat{\varv}_r (r, \ell, m, \nu)|^2 \; .
\end{equation}

\subsection{Impact of the excitation regions on the power spectrum}

The power spectrum of the signal at three depths, $r = [0.243, 0.485, 0.728]$~$R_\star$, expanded on the spherical harmonics with degrees up to $\ell = 100,$ is shown in Fig.~\ref{fig:spectrum_combined_nine_panels_radial_velocity}. For each $\ell$, the power contribution of each azimuthal number, $m$, has been summed quadratically. The spectra from the three models are extremely different, which emphasises that neither the presence of the envelope nor the presence of the core can be omitted in order to model the behaviour of IGWs in the radiative interior.

Internal gravity waves propagate inside the resonant cavity located inside the radiative interior. The location of the top and bottom boundaries of these cavities only depends on the wave frequency and corresponds to the local value of the Brunt-Väisälä frequency, $N$ (see Fig.~\ref{fig:brunt_vaisala}). The IGWs in each spectrum are separated into two distinct regions \citep{Alvan2014,Ahuir2021b,Breton2022simuFstars}, with a boundary materialised by the degree-dependent cut-off frequency represented as the dashed grey lines in Fig.~\ref{fig:spectrum_combined_nine_panels_radial_velocity}. Below this cut-off the progressive IGWs are unable to form standing modes because, excited at one end of the resonant cavity, they are damped by radiative diffusion before reaching the opposite boundary \citep{Zahn1997}. The spectrum of progressive waves is similar between the Full model and the EnvCZ model, while no progressive IGW is visible in the CoreCZ model. This underlines the fact that the convective envelope alone is able to excite these progressives waves for a large range of frequencies and degrees. 

\begin{figure}
    \centering
    \includegraphics[width=0.99\linewidth]{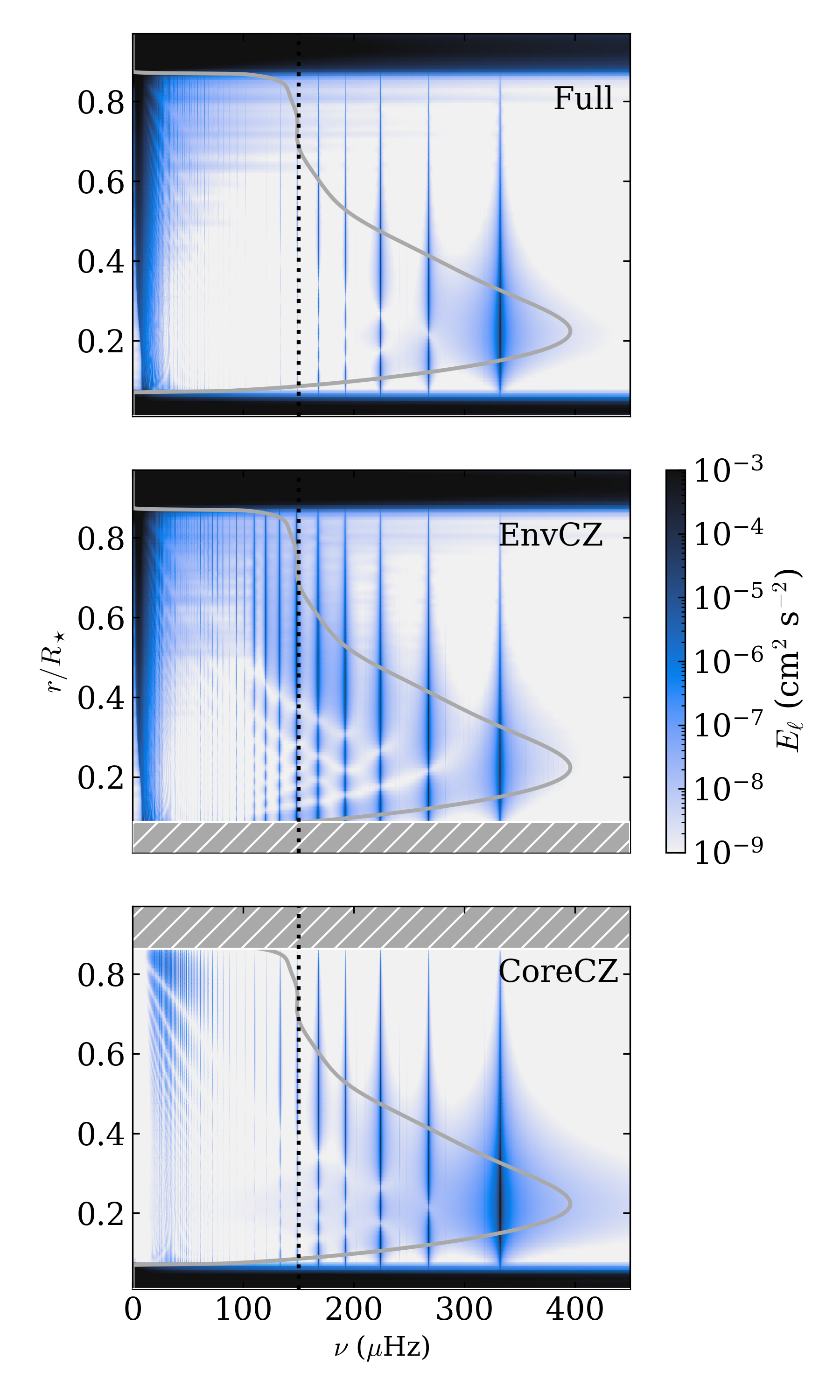}
    \caption{Power spectra $E_\ell$ as a function of depth for $\ell=5$. The thick grey lines correspond to the Brunt-Väisälä frequency, while the dotted vertical black line indicates the 150~$\mu$Hz frequency limit discussed in the text. The hashed areas in grey correspond to the stellar regions that are not included in the EnvCZ and CoreCZ models.}
    \label{fig:spectrum_combined_three_panels_radial_velocity_depth_l5}
\end{figure}

\begin{figure}
    \centering
    \includegraphics[width=0.99\linewidth]{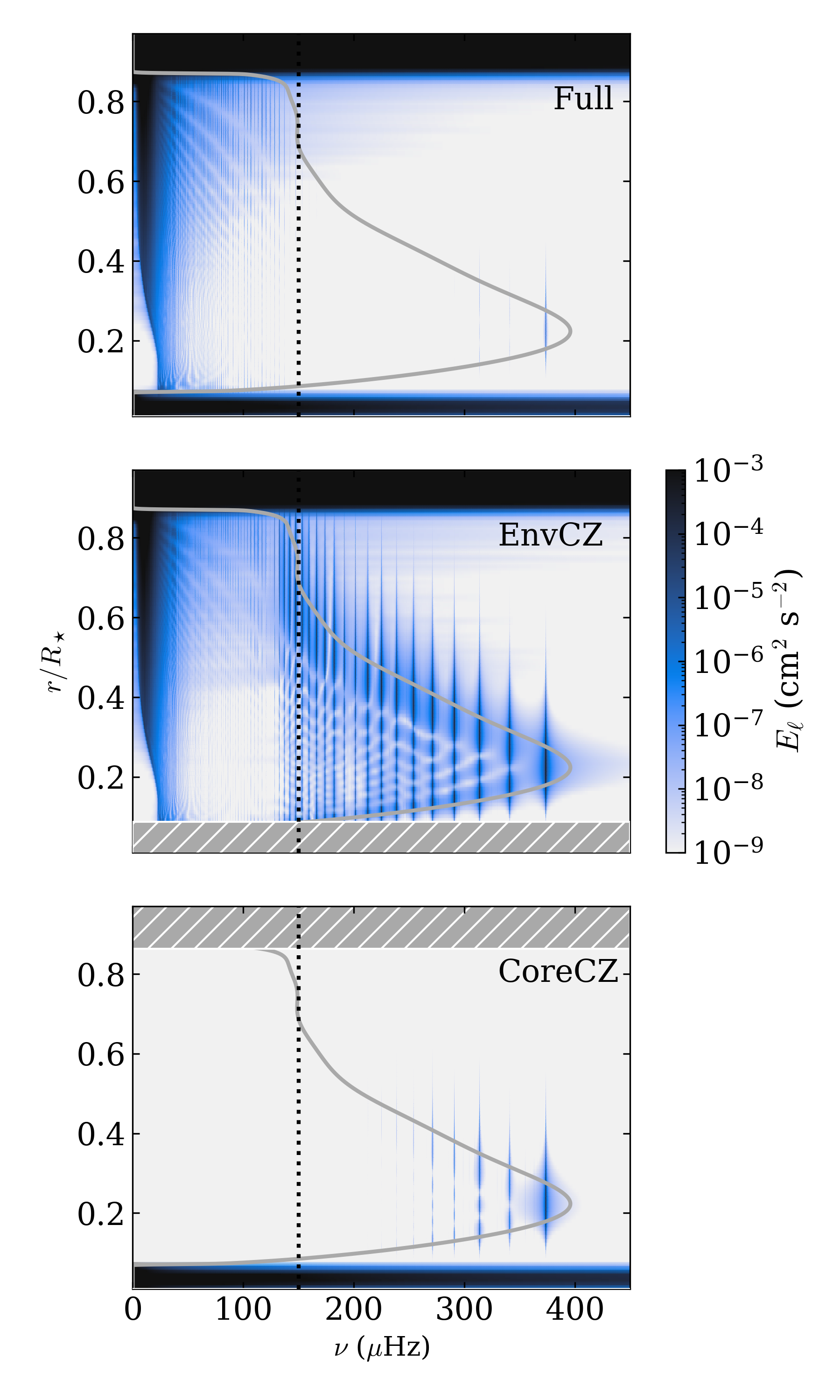}
    \caption{Same as Fig.~\ref{fig:spectrum_combined_three_panels_radial_velocity_depth_l5} for $\ell = 15$.}
    \label{fig:spectrum_combined_three_panels_radial_velocity_depth_l15}
\end{figure}

Above the cut-off frequency, standing modes are visible; launched from one boundary, the IGWs travel through the entirety of the resonant cavity and reach the opposite boundary with sufficient energy. There, a fraction of the wave energy is reflected towards the cavity, allowing the standing mode to form, while another fraction is transmitted outside the cavity as an evanescent tail. This reasoning corresponds to an asymptotic representation of the wave behaviour ($n \gg 1$), where the boundary of the resonant cavity can be validly identified as the radius where $\nu_{n \ell} = N$, with  $\nu_{n \ell}$ the mode frequency. In the following, we see that non-asymptotic modes have a more complex behaviour. Indeed, it is very striking that the rich pattern of high-degree, high-frequency modes exhibited by the EnvCZ model is not present for the Full model. We also note that, at a high frequency and intermediate degrees ($5 \lesssim \ell \lesssim 20$), the CoreCZ model also exhibits a slightly richer mode spectrum. We further illustrate these findings with Figs.~\ref{fig:spectrum_combined_three_panels_radial_velocity_depth_l5} and \ref{fig:spectrum_combined_three_panels_radial_velocity_depth_l15}, where we show the evolution of $E_\ell$ as a function of depth for $\ell=5$ and $\ell=15$, respectively. For $\ell=5$ modes, low-order modes are visible in the three models. We note that the signature of the mode extends beyond the $N = \nu$ boundary. On the contrary, for $\ell = 15$ modes, the mode behaviour above 150~$\mu$Hz is very different from one model to another. As expected from Fig.~\ref{fig:spectrum_combined_nine_panels_radial_velocity}, the EnvCZ model has a rich spectrum of excited eigenmodes above 150~$\mu$Hz, while only a few are visible in the CoreCZ case. Eigenmodes above this frequency are barely visible in the Full case, and they are orders of magnitude below the EnvCZ model in terms of power. Before discussing the condition that allows the modes to form, we emphasise that, despite this depletion in high-degree modes, the Full model exhibits large-amplitude low-degree modes, and these are the ones most susceptible to be detected in disc-resolved observations \citep{Garcia2019}. 

\subsection{Interpreting the convective core's role in mode formation}

As highlighted by the dotted line in Fig.~\ref{fig:spectrum_combined_nine_panels_radial_velocity}, the modes that are unable to form in the Full case, while being excited to large amplitudes in the EnvCZ case, have a frequency $\nu \gtrsim 150 \, \mu$Hz. Below this frequency, the lower boundary of the mode resonant cavity approximately coincides with the interface between the core and the radiative interior (see Fig.~\ref{fig:brunt_vaisala}). We suggest that the distinct behaviour observed for these modes in the Full and EnvCZ cases can be interpreted in terms of mode asymptoticity. We argue that the waves that correspond to an asymptotic mode will perceive  the $\nu_{n \ell} = N$ boundary (in this case, the top of the convective core) as a transmission-reflection interface. It can be demonstrated through the use of the Werner, Jeffreys, Brillouin, and Kramers \citep[WKBJ, e.g.][]{Unno1989} method that their radial eigenfunction is oscillatory within the bounds of the cavity and is thus defined and evanescent outside. Under this asymptotic configuration, in the EnvCZ case, the inner boundary condition imposes complete reflection at $r=0.07 \; R_\star$ (it is the bottom of the simulation domain), while in the Full case a sufficient fraction of the wave energy is reflected and a standing mode is still able to form. 

\begin{figure}[ht!]
    \centering
    \includegraphics[width=0.8\linewidth]{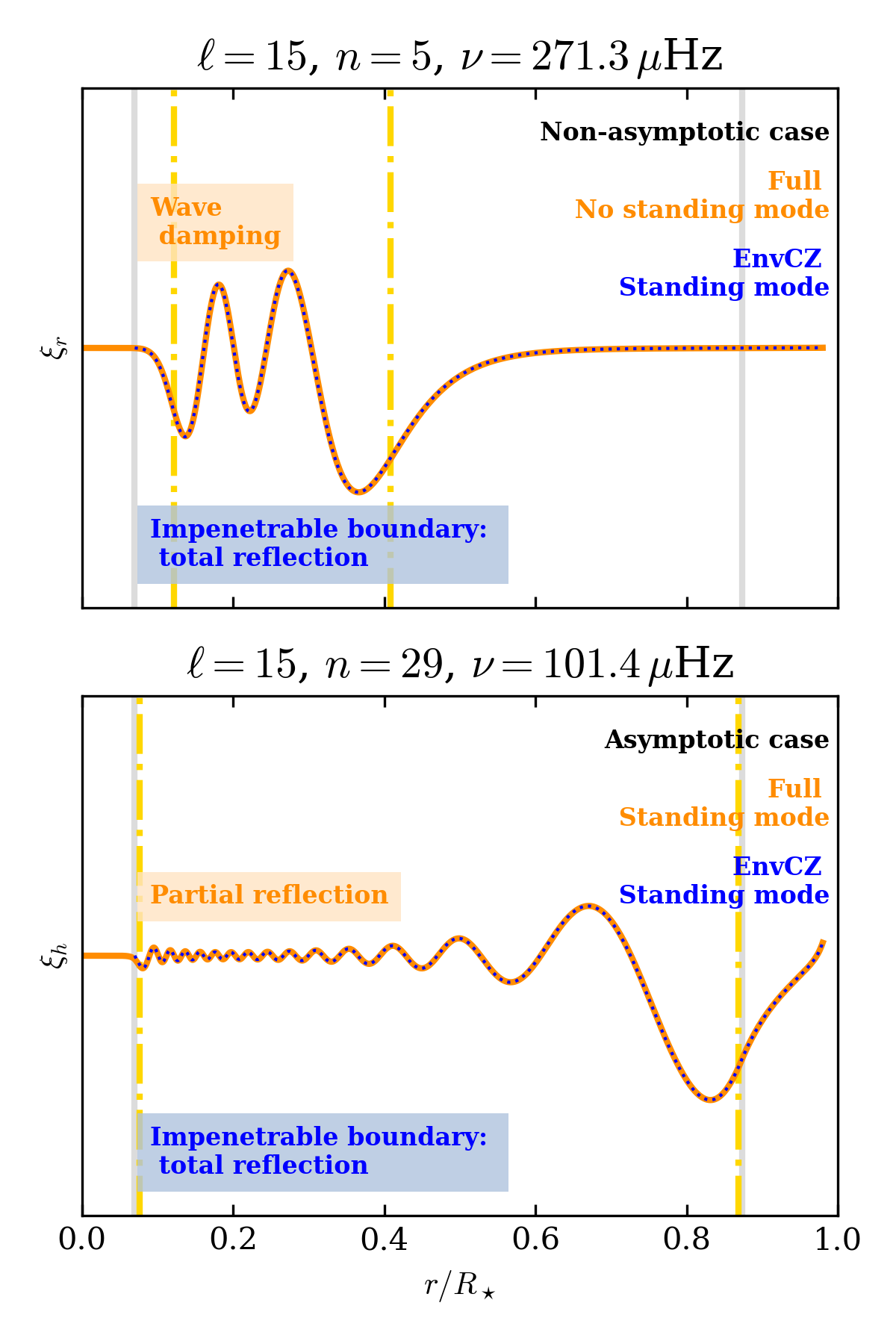}
    \caption{Radial displacement ($\xi_r)$ computed with the Full reference state (orange) and the EnvCZ reference state (dotted blue). The top panel shows a non-asymptotic $\ell=15$, $n=5$ mode with $\nu = 290.8$~$\mu$Hz, while the left panel shows a $\ell=15$, $n=5$ mode with $\nu = 100.1$~$\mu$Hz, which can be interpreted as asymptotic. The vertical dashed grey lines correspond to the bounds of the radiative interior while the vertical dashed yellow lines highlight the radial location of the $\nu_{n \ell} = N$ boundary. On both panels, $\xi_r$ was normalised by its maximal value.}
\label{fig:asymptotic_vs_non_asymptotic}
\end{figure}

By contrast, the WKBJ approximation is no longer valid for non-asymptotic modes. A travelling wave will not perceive the $\nu_{n \ell} = N$ boundary as a transmission-reflection interface and it will propagate beyond this point. In the EnvCZ case, the wave will therefore reach the bottom of the simulation domain and will be reflected there. The situation is different when the dynamical picture includes the convective core. The wave would need to be refracted close to the centre of the star to form a coherent standing mode. However, as it enters the overshoot region, the wave starts experiencing viscous damping from the convective eddies, which adds to the radiative damping. In this picture, the wave is damped while travelling in the convective core, and is therefore unable to form a mode. In the asymptotic picture, the damping efficiency increases with the degree, $\ell$, and the distribution we obtain in the power spectrum for the Full case supports that this should still be valid as only low-degree non-asymptotic modes are able to form. We illustrate our argument with Fig.~\ref{fig:asymptotic_vs_non_asymptotic}, where we compare the eigenfunctions computed with the GYRE code \citep[][see Appendix~\ref{sec:linear_eigenfunction}]{Townsend2013} for the simulation reference states, for two $\ell=15$ modes. The first mode, with $n=5$ and $\nu = 290.8$~$\mu$Hz, is excited in the EnvCZ case but not in the Full case. The second mode, with $n=29$ and $\nu = 100.1$~$\mu$Hz, is excited in both cases. In particular, as visible from the location of the vertical dashed lines, the oscillatory behaviour of the eigenfunction in the asymptotic case is confined inside the whole extent of the radiative interior, while there is no clear distinction between an oscillatory component and an exponential one in the non-asymptotic case. 

\section{Mode detectability \label{sec:mode_detectability}}

We now turn to the issue of mode detectability.
Observationally, the main issue stems from the fact that the low-surface amplitude of $g$ modes makes it difficult to disentangle their contribution from the incoherent turbulent noise that dominates the Fourier spectrum in this range of frequency \citep[e.g.][]{Garcia2007,Breton2023}. In order to examine this issue with our simulation, we developed a process to compute pseudo-observational time series.

\subsection{Observational time series reconstruction}

\begin{figure}[ht!]
    \centering
    \includegraphics[width=0.99\linewidth]{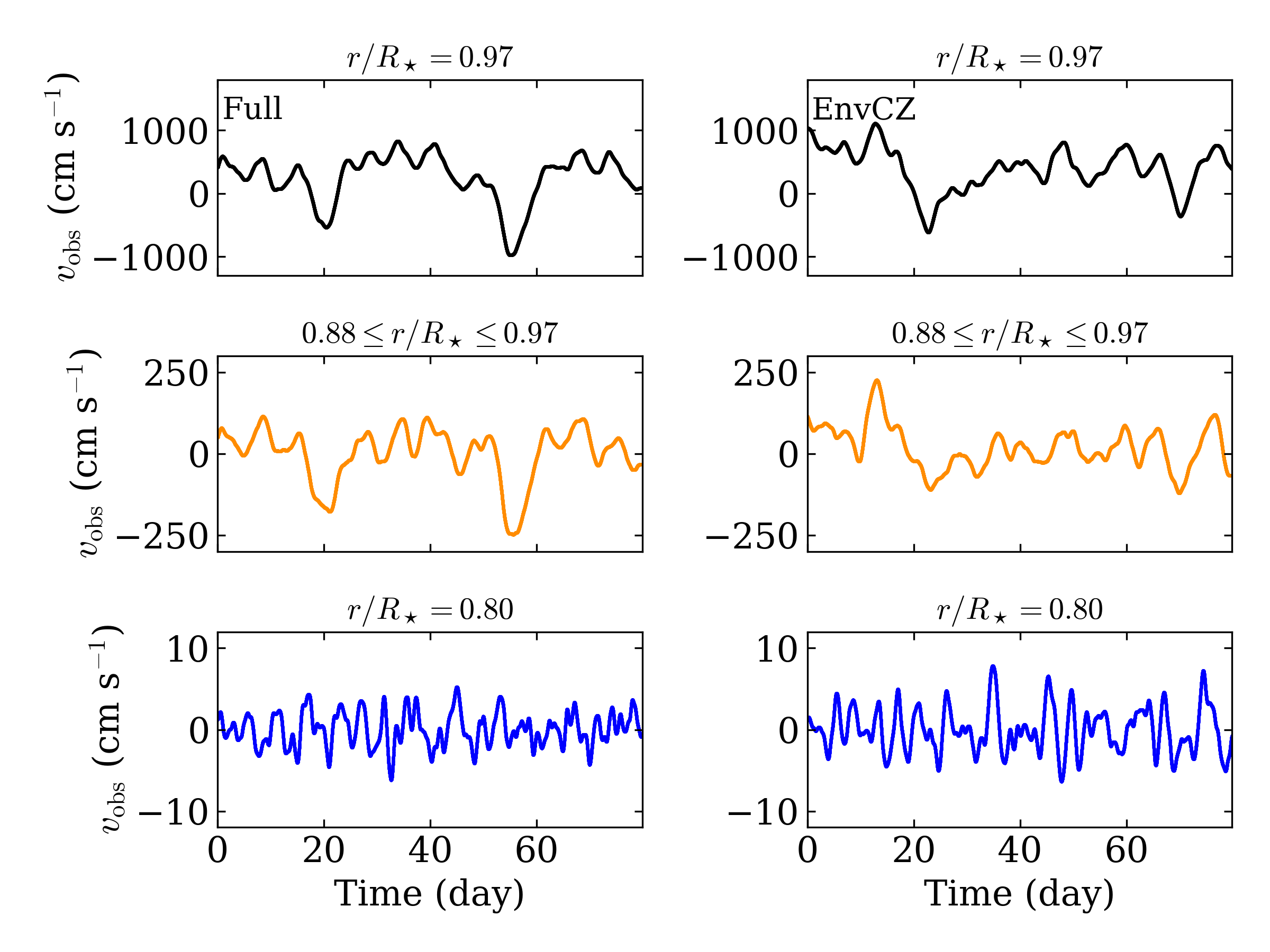}
    \caption{Time series obtained for $\varv_{\rm obs}$ before filtering, at $r/R_\star = 0.97$ (top row), averaged in the convective envelope ($0.88 \leq r/R_\star \leq 0.97$, middle row), and $r/R_\star = 0.80$ (bottom row), for the Full (left column) and the EnvCZ (right column) models, respectively.
    }
    \label{fig:timeseries_comparison_pre_filter}
\end{figure}

\begin{figure}[ht!]
    \centering
    \includegraphics[width=0.99\linewidth]{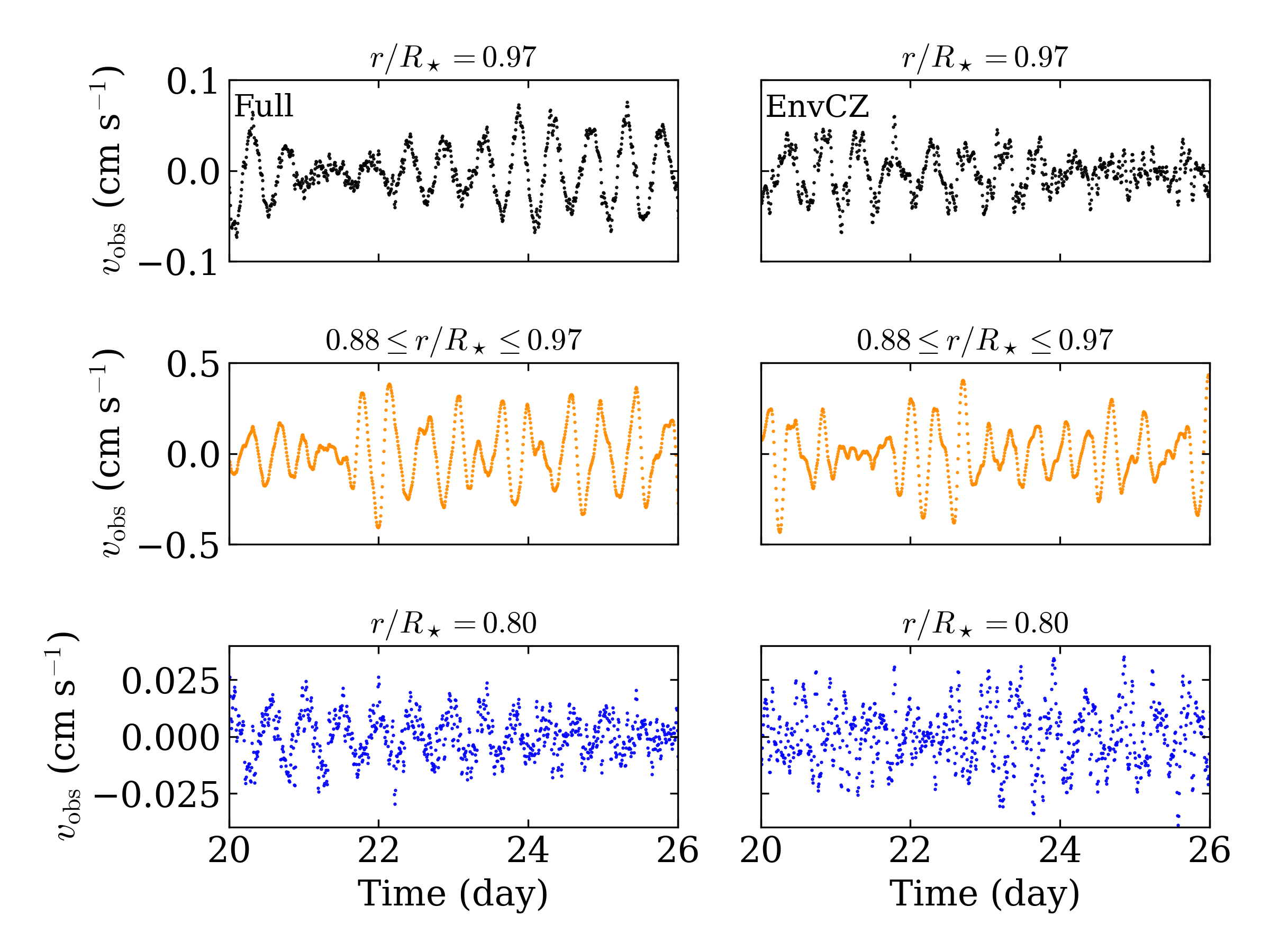}
    \caption{Six-day chunk of the time series obtained for $\varv_{\rm obs}$ after filtering, at $r/R_\star = 0.97$ (top row), averaged in the convective envelope ($0.88 \leq r/R_\star \leq 0.97$, middle row), and $r/R_\star = 0.80$ (bottom row), for the Full (left column) and the EnvCZ (right column) models, respectively.}
    \label{fig:timeseries_comparison_post_filter_zoom_in}
\end{figure}

\begin{figure*}[ht!]
    \sidecaption
    \centering
    \includegraphics[width=12cm]{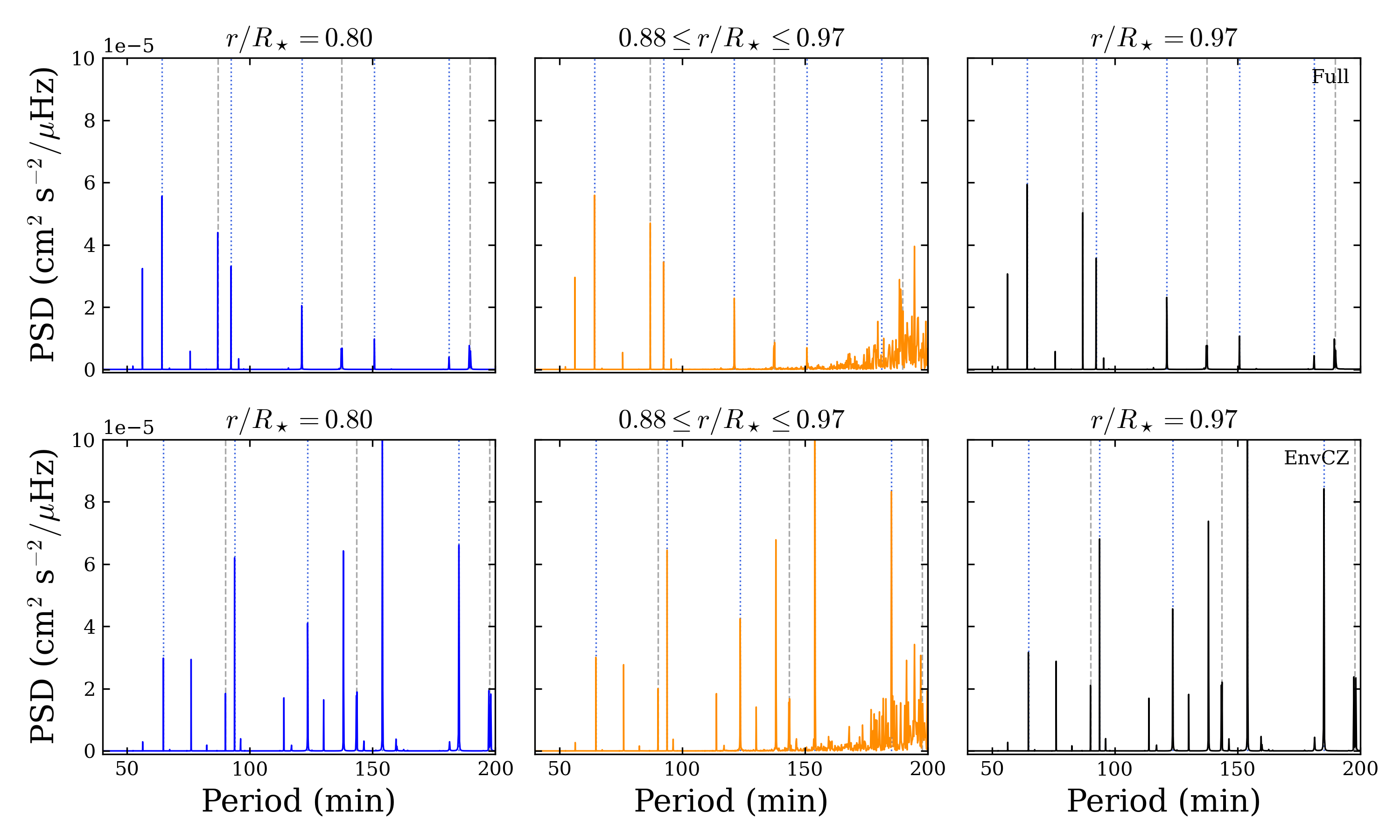}
    \caption{PSD of the disc-integrated projected velocity signal, $\varv_{\rm obs}$, obtained from the simulation in the upper radiative interior ($r/R_\star = 0.80$, \textit{left column}), averaged in the convective envelope ($0.88 \leq r/R_\star \leq 0.97$, \textit{middle column}), and close to the top of the simulation domain ($r/R_\star = 0.97$, \textit{right column}). The PSD obtained for the Full model is shown at the top and the EnvCZ model PSD is at the bottom. The regular pattern of the $\ell=1$ modes is highlighted with vertical dashed grey lines and that of the $\ell=2$ modes with the vertical dotted blue lines.}
    \label{fig:integrated_timeseries}
\end{figure*}

In order to produce the pseudo-observational power spectral densities (PSD), we applied the following process to our data. We denote $\theta_0$ and $\phi_0$ as the co-latitude and the longitude corresponding to the observer line of sight. The resultant projected velocity, $\varv_{\rm proj}$, in this direction is therefore the sum of the three components $\varv_{r, {\rm proj}}$, $\varv_{\theta, {\rm proj}}$, and $\varv_{\phi, {\rm proj}}$
\begin{equation}
    \varv_{\rm proj} (\theta, \phi) = \varv_{r, {\rm proj}} (\theta, \phi) + \varv_{\theta, {\rm proj}} (\theta, \phi) + \varv_{\phi, {\rm proj}} (\theta, \phi) \; ,
\end{equation}
where
\begin{align}
    \varv_{r, {\rm proj}} (\theta, \phi) &= \varv_r \cos (\theta - \theta_0) \cos (\phi - \phi_0) \; ,\\
    \varv_{\theta, {\rm proj}} (\theta, \phi) &= \varv_\theta \sin (\theta_0 - \theta) \cos (\phi - \phi_0) \; , \\
    \varv_{\phi, {\rm proj}} (\theta, \phi) &= \varv_\phi \sin (\phi_0 - \phi) \sin \theta_0 \; ,
\end{align}
where $\varv_\theta$ and $\varv_\phi$ are the latitudinal and longitudinal component of the fluid velocity, respectively. The angle $\psi$ between the normal vector of a given surface element of the sphere and the observer line of sight follows the relation, denoting $\mu = \cos \psi$ \citep{Lanza2016,Breton2024}
\begin{equation}
    \mu (\theta, \phi) = \sin \theta_0 \sin \theta \cos (\phi-\phi_0) + \cos \theta_0 \cos \theta \; ,
\end{equation}
and only the elements with $\mu > 0$ are visible to the observer. In what follows, we consider for simplicity $\phi_0 = 0$ and $\theta_0 = \pi/2$ in the co-rotating frame. This means that our fictitious observer is rotating around the star with the same angular velocity. Nevertheless, the advantage of this approach is that the same surface elements continually remain in the observer's field of view, and we do not introduce a discrete spurious modulation related to the appearance and disappearance of surface elements when they cross the $\mu = 0$ threshold.  
We model the limb-darkening contribution, $\mathcal{L}_d$, with the Eddington law \citep{Garcia2019} so that 
\begin{equation}
    \mathcal{L}_d (\mu) = (0.4 + 0.6 \mu) \mu \; .
\end{equation}

Finally, the observed velocity, $\varv_{\rm obs}$ , is written 
\begin{equation}
    \varv_{\rm obs} = \frac {\displaystyle\sum\limits_{i, j} a_i \varv_{\rm proj} (\theta_i, \phi_j) \mathcal{L}_d (\mu_{i,j}) \mu_{i,j}} {\displaystyle\sum\limits_{i, j} a_i \mu_{i,j}} \; ,
\end{equation}
where the indices $(i,j)$ identify each surface element and $a_{i} = \sin \theta_{i}$ is the corresponding relative surface. {In what follows,} we reconstruct $v_{\rm obs}$ considering angular degrees $\ell$ from {0 to 100, which is up to the largest angular degree we stored for the time series}. {Nevertheless, the additional tests we performed showed that the reconstruction did not significantly differ considering a truncation at a significantly lower $\ell$, such as $\ell = 10$.}

{In our simulations, the convective noise contribution at a high frequency mostly results  from a power leakage of the low-frequency signal associated with convection. This means that, by applying a high-pass filter on our reconstructed time series, we are able to increase the signal-to-noise ratio of the periodic signatures at a higher frequency, which corresponds, in our case, to the $g$-mode signal.}

In order to remove the contamination from the convective noise in the high-frequency range of the PSD, a high-pass finite-impulse-response (FIR) filter with a {12-hour ($\sim$23~$\mu$Hz) cut-off is therefore} applied on the time series \citep{Breton2022,Breton2024Plato} before computing the Fast Fourier Transform (FFT). In Fig.~\ref{fig:timeseries_comparison_pre_filter}, we show the time series we obtain before applying the FIR filter. As shown, {due to the contribution of the convective motions}, the amplitude of the $\varv_{\rm obs}$ signal significantly varies as a function of the depth, with a maximum amplitude of about $\num{1e3}$~cm~s$^{-1}$ at $r/R_\star = 0.97$ and only 10~cm~s$^{-1}$ at $r/R_\star = 0.80$, in the radiative interior. It also clearly appears  that the long-period variations of the time series at $r/R_\star = 0.97$ are strongly correlated with the average variation in the convective envelope, which reflects the impact of the large-scale structures that are visible in Fig.~\ref{fig:volume_display}. In Fig.~\ref{fig:timeseries_comparison_post_filter_zoom_in}, we show a six-day chunk of the same time series after applying the 12-hour cut-off FIR filter. With the long-period trend removed, the typical amplitude of the signal is now much smaller. 
As can be seen by comparing the three rows of the figure, some contribution from the convection in the period range shorter that 12 hours remains, and enhances the total amplitude of the signal observed in the convective envelope with respect to the radiative interior. 

\subsection{Observational PSD}

By integrating the projected velocity signal on the stellar disc, as would be seen by an observer, we were therefore able to construct a set of time series that allows us to model how the disc-integrated observable $g$-mode signature evolves with the stellar depth. In Fig~\ref{fig:integrated_timeseries}, we show the corresponding PSD obtained after applying a DFT on these time series. It is expected that, in this observational PSD, the contribution from increasing degree $\ell$ will be gradually filtered out by the disc-integrated spatial integration \citep{Dziembowski1977}. Moreover, by temporally filtering the low-frequency contribution to the time series, we are able to unambiguously isolate the $g$-mode signal in the PSD, even in the upper layers of the convective envelope. Here again, the signature differs importantly between the Full case and the EnvCZ model. In both cases, the main $g$-mode signature contribution comes from the $\ell=1$ and $\ell=2$ modes. As visible, the EnvCZ case exhibits a few peaks related to the contribution from modes at higher degrees (this is also the case for the mode with the lowest frequency in the Full spectrum). The existence of this signature demonstrates without ambiguity that the evanescent tail of the $g$ modes maintains its coherence up to the top of the convective envelope and should be able to reach the stellar photosphere. Removing the incoherent contribution from convective noise allows us to reveal the presence of the $g$ -mode pattern, regularly spaced in periods {for each degree~$\ell$} \citep{Tassoul1980}. The convective core has an imprint not only on this period spacing, but also by limiting the number of $g$ modes. In particular, our results highlight the need to look for the combined signature of $\ell=1$ and $\ell=2$ modes in observational data, as they should reach the surface of the star with comparable amplitude.

\section{Conclusion \label{sec:conclusion}}

In this work, we presented the first 3D simulations of the dynamics of a solar-type star with two convective zones. F-type stars possess both an external convective envelope and a small convective core that surrounds a stably stratified radiative interior. In order to investigate the respective contribution from the core and the envelope in the excitation of $g$ modes, we compared a full-extent model (Full) with a coreless model (EnvCZ) and a model with no convective envelope model (CoreCZ). We also highlighted the crucial role that the convective core plays regarding the excitation of the modes, as it prevents high-degree modes from forming in the non-asymptotic high-frequency range. By integrating the signal on the stellar disc in order to compute pseudo-observational time series, we demonstrated that, despite their evanescence, $g$ modes maintain their coherent integrity up to the top of the convective turbulent layer, which supports the claim that they have a detectable observable signature at the stellar photosphere. 

This work paves the way for a systematic search for $g$ -modes signatures in the F-type solar-like pulsators that will be observed by the PLATO mission. A first attempt at such an analysis was performed by \citet{Breton2023} considering the 34 F-type solar pulsators with the best characterisation in the \textit{Kepler} mission \citep{Borucki2010,Mathur2014,Davies2016,Lund2017,SilvaAguirre2017}, but definitive conclusions could not be drawn given the limited size of the sample.
Given the low amplitude of the modes, they will have to be looked for in a large sample of stars and using analysis techniques built on solid statistical ground.   

\begin{acknowledgements}
The authors thank the anonymous referee for comments and suggestions that helped improving the manuscript. S.N.B acknowledges support from PLATO ASI-INAF agreement no. 2022-28-HH.0 "PLATO Fase D". SNB acknowledges support from the INAF grant MASTODINT. A.S.B acknowledges support from Solar Orbiter CNES grant and financial support by ERC Whole Sun Synergy grant \#810218. While partially funded by the European Union, views and opinions expressed are, however, those of the authors only and do not necessarily reflect those of the European Union or the European Research Council. Neither the European Union nor the granting authority can be held responsible for them.
A.S.B acknowledges CNES support for PLATO WP123400. R.A.G acknowledges support from PLATO and GOLF CNES grants. 

\end{acknowledgements}

\bibliographystyle{aa} 
\bibliography{biblio.bib} 

\appendix
\section{Hydrodynamical setup \label{sec:numerical_setup}}

The ASH numerical setups presented in Sect.~\ref{sec:simulation_setups} are defined in the spherical co-rotating frame $(r, \theta, \phi)$, with unit vectors $(\bm{e}_r, \bm{e}_\theta, \bm{e}_\phi)$. The reference density, pressure, temperature, and specific entropy are denoted as $\bar{\rho}$, $\bar{P}$, $\bar{T}$, and $\bar{S}$, with corresponding fluctuations $\rho$, $P$, $T$, $S$. They are connected through the equation of state and the zeroth-order ideal gas law
\begin{align}
    \label{eq:eq_of_state}
    \frac{\rho}{\bar{\rho}} &=  \frac{P}{\bar{P}} -  \frac{T}{\bar{T}}
    =  \frac{P}{\gamma \bar{P}} - \frac{S}{c_p} \; , \\
    \bar{P} &= \mathcal{R} \bar{\rho} \bar{T} \; ,
\end{align}
where $c_p$ is the specific heat per unit mass at constant pressure, $\gamma$ the adiabatic exponent, and $R$ is the gas constant. The quantities $S$ and $c_p$ can be combined with the gravitational acceleration $\bm{g} = - g \bm{e}_r$ to define the Brunt-Väisälä frequency, $N$, through the relation
\begin{align}
    N^2 = \frac{g}{c_p} \diff{S}{r} \; .
\end{align}
In the WKBJ approximation, $N >0$ defines the resonant cavities of IGWs. It is also directly related to the asymptotic period spacing of consecutive $g$ modes of same degree, $\Delta P_\ell$, through \citep{Tassoul1980}
\begin{equation}
    (\Delta P_\ell )^{-1} = \frac{\sqrt{\ell (\ell+1)}}{2 \pi^2} \int_{r_1}^{r_2} \frac{N}{r} \mathrm{d}r \; ,
\end{equation}
where $r_1$ and $r_2$ correspond to the boundaries $N^2 =0$. $N$ is represented in Fig.~\ref{fig:brunt_vaisala}. We note that the maximal value of $N$ is reached in the deep radiative interior, close to $r/R_\star = 0.25$. 

\begin{figure}[ht!]
    \centering
    \includegraphics[width=0.99\linewidth]{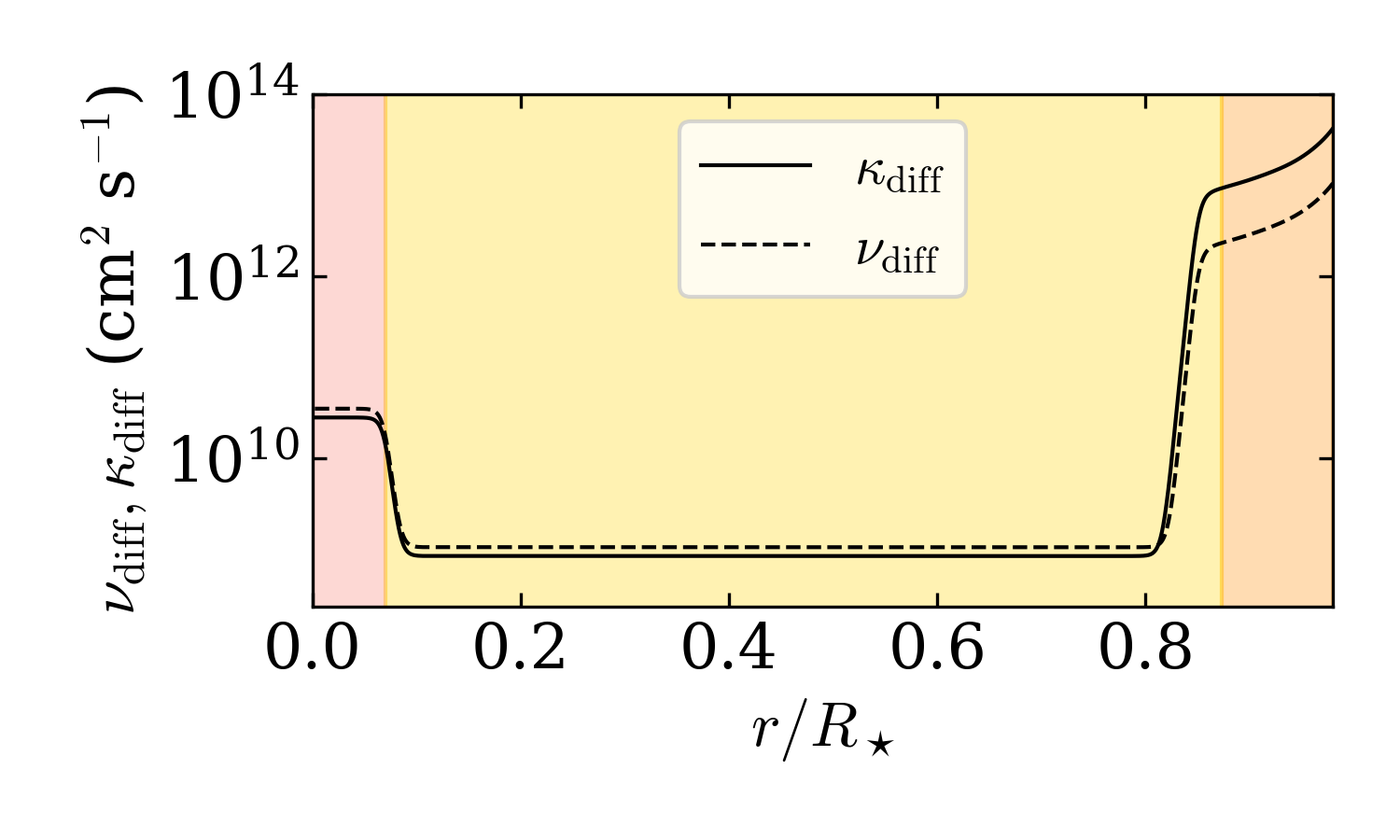}
    \caption{$\nu_{\rm diff}$ (solid line) and $\kappa_{\rm diff}$ (dashed line) diffusivity profiles used in the simulations. The extent of the convective core, the radiative interior, and the convective envelope are indicated in colour as in Fig.~\ref{fig:brunt_vaisala}.}
    \label{fig:nu_kappa_diffusivities}
\end{figure}

The hydrodynamic equations are solved in the  Lantz-Braginsky-Roberts \citep[LBR,][]{Lantz1992,Braginsky1995} formulation of the anelastic approximation \citep{Gough1969}. By filtering out acoustic waves, the anelastic approximation allows for larger integration time steps with respect to fully compressible setups. The LBR formulation is specifically chosen because it was shown that it preserves better the IGWs energy \citep{Brown2012}. The momentum equation is
\begin{equation}
\label{eq:momentum}
\bar{\rho} \left( \diffp{\bm{\varv}}{t} + ({\bm{\varv} \cdot \nabla}) \bm{\varv} \right)
= - \bar{\rho} \nabla \tilde{\omega} - \bar{\rho} \frac{S}{c_p} \bm{g} - 2\bar{\rho}\bm{\Omega}_0 \times \bm{\varv} - \nabla \cdot \bm{\mathcal{D}} \; ,
\end{equation}
where $\tilde{\omega} = P/\bar{\rho}$ is the reduced pressure fluctuation, $\bm{\varv} = (\varv_r, \varv_\theta, \varv_\phi)$ is the local velocity, $\bm{\Omega}_0 = \Omega_0 \bm{e}_z$ is the rotation frequency of the reference frame. The viscous stress tensor, $\bm{\mathcal{D}}$, is
\begin{equation}
    \label{eq:reynold_tensor}
    \mathcal{D}_{ij} = - 2 \bar{\rho} \nu_{\rm diff} \left( e_{ij} - \frac{1}{3}(\nabla \cdot \bm{\varv}) \delta_{ij} \right) \; ,
\end{equation}
with $\nu_{\rm diff}$ the effective viscous diffusivity, $e_{ij} = 1/2 \left( \partial_j \varv_i + \partial_i \varv_j \right)$ the strain rate tensor, and $\delta_{ij}$ the Kronecker symbol. In the anelastic approximation, the continuity equation is expressed as:
\begin{equation}
    \nabla \cdot (\bar{\rho} \bm{\varv}) = 0 \; .
\end{equation}
Our equation of conservation of internal energy writes
\begin{equation}
\begin{split}
    \label{eq:conservation_energy}
    \bar{\rho} \bar{T} \diffp{S}{t} + \bar{\rho}\bar{T}\bm{\varv} \cdot \nabla \Big( S + \bar{S} \Big) &= \\
    \bar{\rho}\epsilon + \nabla \cdot \Bigg[ \kappa_r\bar{\rho} c_p \nabla \Big( T + \bar{T} \Big)
    &+ \kappa_{\rm diff} \bar{\rho} \bar{T} \nabla S + \kappa_0 \bar{\rho} \bar{T} \nabla \bar{S}  \Bigg] \\
    &+ 2 \bar{\rho} \nu_{\rm diff} \left[ e_{ij}e_{ij} - \frac{1}{3} (\nabla \cdot \bm{\varv})^2 \right] \; ,
\end{split}
\end{equation}
where $\kappa_r$ is the radiative diffusivity \citep{Kippenhahn2012}. The $\epsilon$ term accounts for the energy generation by nuclear burning, modelled as a temperature power-law, $\epsilon = \epsilon_0 \bar{T}^k$. Similarly to $\nu_{\rm diff}$, $\kappa_{\rm diff}$ is an effective diffusivity while the diffusivity $\kappa_0$ is set to carry the unresolved entropy eddy flux in the convective zone near the surface \citep[see][]{Brun2004}. We set $\kappa_{\rm diff}$ and $\nu_{\rm diff}$ as a trade-off between the desired turbulent regime and numerical tractability. In particular, we decrease both $\kappa_{\rm diff}$ and $\nu_{\rm diff}$ in the radiative interior in order to minimise the IGWs damping. The $\nu_{\rm diff}$ and $\kappa_{\rm diff}$ are described by an equation of the form
\begin{equation}
\begin{split}
    \nu_{\rm diff} = \nu_1 \Bigg[ \beta + \frac{1 -\beta}{2} \left( \frac{\bar{\rho}_\mathrm{top}}{\bar{\rho}}  \right)^\frac{1}{2} \left(\tanh \frac{r-r_{t1}}{\sigma_t} + 1\right) \Bigg] \\
    + \frac{\nu_2}{2} \left(\tanh \frac{r-r_{t2}}{\sigma_t} + 1 \right)  \; .
\end{split}
\end{equation}
with, in both case, the transition radii $r_{t1} = \num{8.9e10}$~cm and $r_{t2} = \num{7.3e9}$~cm, the stiffness $\sigma_t = \num{8e8}$~cm. This way, at the top of the simulation domain $\nu_{\rm diff, top} \approx \nu_1$, in the radiative interior, $\nu_{\rm diff, radiative} \approx \beta \nu_1 $, and in the convective core, $\nu_{\rm diff, core} \approx \nu_2 + \beta\nu_1$. We use $\nu_1 = \num{1.05e13}$~cm$^2$~s$^{-1}$, $\nu_2 = \num{3.395e10}$~cm$^2$~s$^{-1}$, $\kappa_1 = \num{4.2e13}$~cm$^2$~s$^{-1}$, and $\kappa_2 = \num{2.716e10}$~cm$^2$~s$^{-1}$. For $\nu_{\rm diff}$, $\beta = \num{1e-4}$, and for $\kappa_{\rm diff}$, $\beta = \num{2e-5}$. Both profiles are represented in Fig.~\ref{fig:nu_kappa_diffusivities}.

We finally have to define the boundary conditions. For the Full and the CoreCZ models, the singularity at $r=0$ is treated with a regularisation procedure that was previously validated for the study of IGWs \citep{Alvan2014}. For every other boundary, we adopt the following conditions: rigid, stress-free impenetrable boundaries with constant mean-entropy gradient \citep{Breton2022simuFstars}.

\begin{figure}[ht!]
    \centering
    \includegraphics[width=0.99\linewidth]{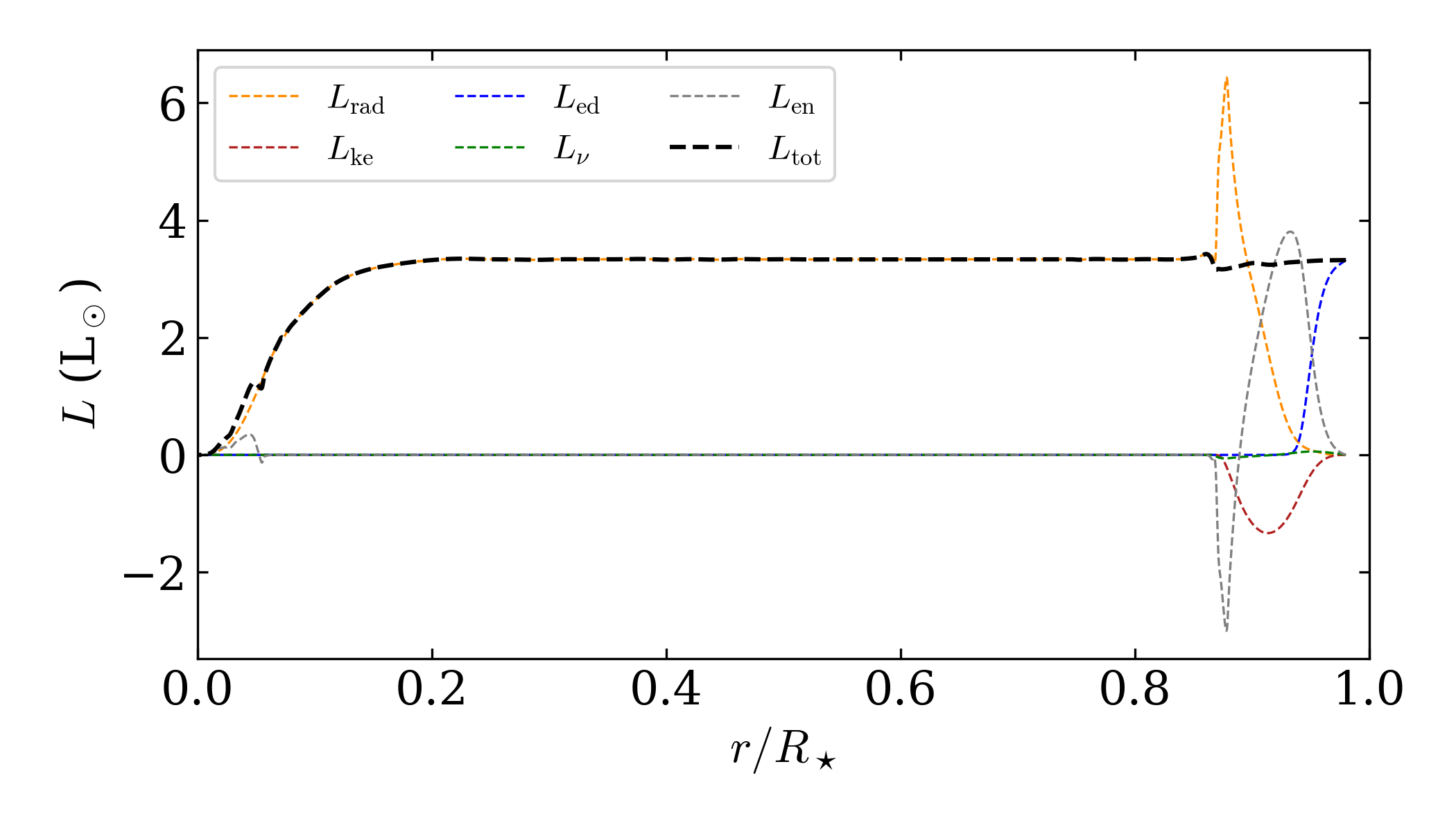}
    \caption{Flux balance obtained for the Full setup. $L_{\rm rad}$ (dashed orange), $L_{\rm ke}$ (dashed red), $L_{\rm ed}$ (dashed blue),  $L_{\rm \nu}$ (dashed green), $L_{\rm en}$ (dashed grey), and $L_{\rm tot}$ are shown. See Eq.~(\ref{eq:flux_balance}) and subsequent explanations.}
    \label{fig:flux_balance}
\end{figure}

\begin{figure}[ht!]
    \centering
    \includegraphics[width=0.99\linewidth]{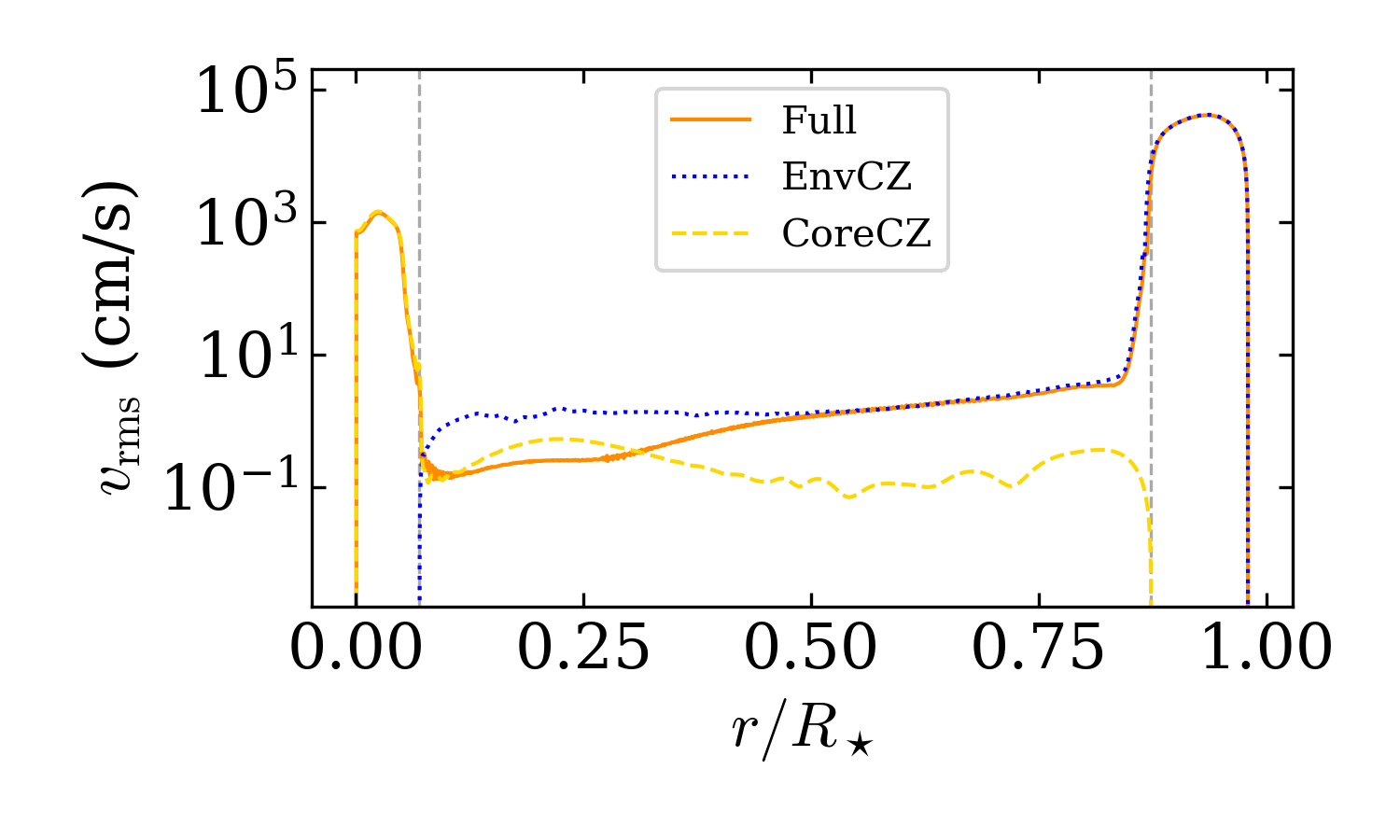}
    \caption{Root-mean-square (rms) radial velocity profile, $\varv_{\rm rms}$, when the steady state is reached for each of the three models. The Full profile is shown in orange, the EnvCZ one in dotted blue, and the CoreCZ one in dashed yellow. The dashed vertical grey lines indicate the boundaries of the convective core and of the convective envelope.}
    \label{fig:v_rms_comparison}
\end{figure}

From the problem formulation, the total energy balance can be expressed as the sum of the flux related to the different transport processes \citep{Brun2004}
\begin{equation}
\label{eq:flux_balance}
    L_{\rm tot} = L_{\rm rad}+L_{\rm ke}+L_{\nu}+L_{\rm en} + L_{\rm ed} \; ,
\end{equation}
where $L_{\rm rad}$ is the radiative luminosity, $L_{\rm ke}$ the kinetic energy luminosity, $L_{\rm \nu}$ the diffusive processes energy luminosity, $L_{\rm en}$ the enthalpy luminosity, and $L_{\rm ed}$ the unresolved eddy luminosity. As explained in the main body of the text, small perturbations introduced in the convective unstable layer at the start of the simulation allow for the convective instability to grow and build the $L_{\rm ke}$ luminosity. The simulation is evolved for several times the convective turnover time, $\tau_{\rm CZ}$, in order to allow it to reach a dynamical steady state. In the convective envelope, $\tau_{\rm CZ} \approx 4$~days while in the core, $\tau_{\rm CZ} \approx 160$~days. We therefore evolve the Full and CoreCZ simulations for about 1250 days in order to reach the dynamical steady state.  In Fig.~\ref{fig:flux_balance}, we show the flux balance achieved by the Full model once it has reached this steady state and is relaxed. As the thermal relaxation timescale \citep{Zahn1991} of the background state (about $\num{1e5}$~yr) is beyond reach, we adjust $L_{\rm rad}$ in the overshoot region to maintain the flux balance \citep{Miesch2000,Breton2022simuFstars}. We also show in Fig.~\ref{fig:v_rms_comparison}, the rms velocity regime that develop in the relaxed simulations for the three cases. Finally, the grid resolution of each model is indicated in Table~\ref{tab:grid_resolution}, with the number of radial ($N_r$), latitudinal ($N_\theta$), and azimuthal ($N_\phi$) mesh points.

\begin{table}[h!]
    \centering
    \caption{Grid resolution of the models.}
    \begin{tabular}{cccc}
         Model & Full & EnvCZ & CoreCZ \\
         \hline
         $N_r$ & 2620 & 1984 & 2157 \\
         $N_\theta$ & 1024 & 1024 & 512 \\
         $N_\phi$ & 2048 & 2048 & 1024 \\
    \end{tabular}
    \label{tab:grid_resolution}
\end{table}

\section{Power spectrum comparison with linear eigenfunctions \label{sec:linear_eigenfunction}}

The linear eigenfunction shown in Fig.~\ref{fig:asymptotic_vs_non_asymptotic} were computed with the GYRE code \citep{Townsend2013}, considering the 1D reference profile of our 3D setup. The implemented GYRE formalism closest to the anelastic approximation used in ASH is the $\gamma$ formalism \citep{Ong2020}, as it isolates the $g$-mode buoyancy cavity from the acoustic cavity, similarly to the anelastic approximation where acoustic waves are filtered out. The oscillation equations are solved both with and without the Cowling approximation, which neglects the perturbation of the gravitational potential \citep{Cowling1941}. 

\begin{figure}[ht!]
    \centering
    \includegraphics[width=0.85\linewidth]{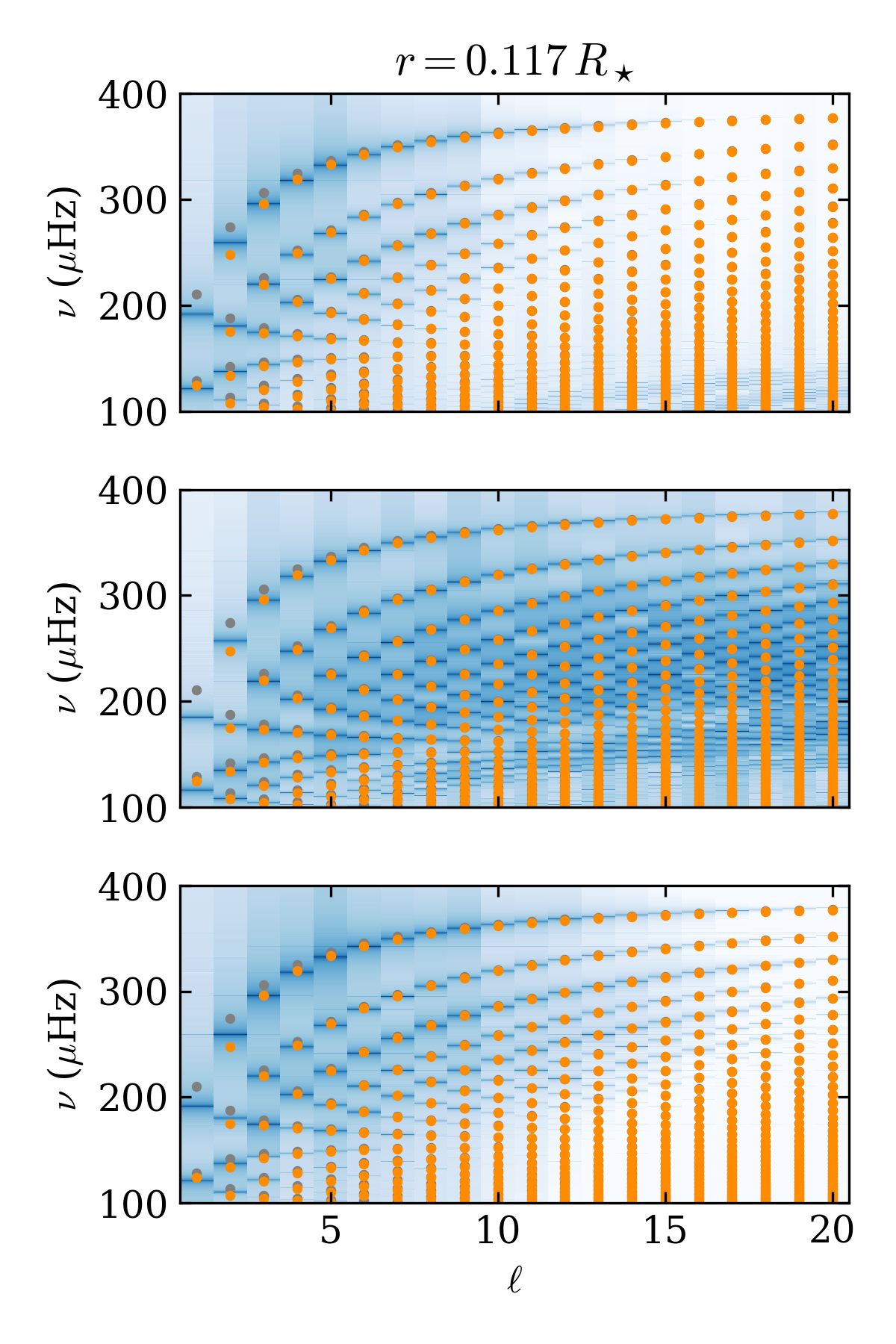}
    \caption{Power spectrum for the Full (top), the EnvCZ (middle), and the CoreCZ (bottom) cases at $r = 0.117 \, R_\star$. The eigenfrequencies computed with GYRE are shown as orange dots for the computations without the Cowling approximation, and as grey dots for the computation with the Cowling approximation. At high degree and high frequency, grey and orange dots are superposed.}
    \label{fig:gyre_ash_comparison}
\end{figure}

In Fig.~\ref{fig:gyre_ash_comparison}, we compare the GYRE frequencies for modes with $\nu_{n \ell} > 100$~$\mu$Hz and $\ell \leq 20$ to the $E_\ell$ distribution computed with the ASH simulations.
As it is visible in Fig.~\ref{fig:gyre_ash_comparison}, the agreement between the GYRE eigenfunctions and the 3D-simulation power spectrum is very good. We note that the first $\ell = 1$ mode of the ASH spectra, visible at $\nu_{n \ell} \approx 190$~$\mu$Hz, is present in the GYRE predictions including the Cowling approximation, but absent otherwise. It corresponds to a $n=0$ dipolar mode that is unphysical for a self-gravitating spherical object (that would mean the object is oscillating around its centre of mass) but permitted in the context of the Cowling approximation and in ASH setups, where the gravitational potential is defined by a fixed background term with neglected fluctuations.

\end{document}